\documentclass[preprint,aps,pra,showpacs,floatfix]{revtex4-1}
\usepackage{amsmath, amssymb}
\usepackage{times}
\usepackage {euscript}
\usepackage{graphicx}
\usepackage{amsthm}
\usepackage{epsfig}
\usepackage{bm}
\usepackage{multirow}
\usepackage{longtable}
\usepackage[dvipsnames]{xcolor}
\newcommand{\bp}{{\bf p}}
\newcommand{\bk}{{\bf k}}
\newcommand{\br}{{\bf r}}
\newcommand{\bb}{{\bf b}}

\newcommand{\bn}{{\bf n}}

\newcommand{\bJ}{{\bf J}}
\newcommand{\balpha}{\boldsymbol{\alpha}}
\newcommand{\bepsilon}{\boldsymbol{\epsilon}}

\newcommand{\tam}{\left(\bJ\cdot \hat{\bn}_0\right)}
\newcommand{\mom}{\left({\bf p}\cdot \hat{\bn}_0\right)}
\graphicspath{{graph/}}
\begin{document}
\thispagestyle{empty}
\title{
Analysis of angular momentum properties of photons emitted in 
fundamental atomic processes
}
\author{V.~A.~Zaytsev$^{1, 2}$,
        A.~S.~Surzhykov$^{3, 4}$,
        V.~M.~Shabaev$^{1}$, and
        Th. St\"ohlker$^{5, 6, 7}$}
\affiliation{
$^1$ Department of Physics, St. Petersburg State University,
     Ulianovskaya 1, Petrodvorets, 198504 St. Petersburg, Russia \\
$^2$ ITMO University, 
     Kronverkskii ave 49, 197101 Saint Petersburg, Russia         \\
$^3$ Physikalisch-Technische Bundesanstalt, 
     D-38116 Braunschweig, Germany \\
$^4$ Technische Universit\"at Braunschweig,
     D-38106 Braunschweig, Germany \\
$^5$ GSI Helmholtzzentrum f\"ur Schwerionenforschung GmbH, 
     D-64291 Darmstadt, Germany \\
$^6$ Helmholtz-Institute Jena, 
     D-07743 Jena, Germany \\
$^7$ Institut f\"ur Optik und Quantenelektronik, 
Friedrich-Schiller-Universit\"at, D-07743 Jena, Germany
\vspace{10mm}
}
%
\begin{abstract}
Many atomic processes result in the emission of photons.
Analysis of the properties of emitted photons, such as energy and 
angular distribution as well as polarization, is regarded as a powerful 
tool for gaining more insight into the physics of corresponding 
processes. 
Another characteristic of light is the projection of its angular 
momentum upon propagation direction.
This property has attracted a special attention over 
the last decades due to studies of twisted (or vortex) light beams.
Measurements being sensitive to this projection may provide valuable 
information about the role of angular momentum in the fundamental atomic 
processes.
Here we describe a simple theoretical method for determination of the 
angular momentum properties of the photons emitted in various atomic 
processes.
This method is based on the evaluation of expectation value of the 
total angular momentum projection operator.
To illustrate the method, we apply it to the text-book examples of 
plane-wave, spherical-wave, and Bessel light. 
Moreover, we investigate the projection of angular momentum for the 
photons emitted in the process of the radiative recombination with ionic 
targets.
It is found that the recombination photons do carry a non-zero projection 
of the orbital angular momentum.
\end{abstract}
%
\maketitle
%
%
%
%
%
%
\section{INTRODUCTION}
In recent decades various experimental techniques have been developed to 
produce beams of light carrying a non-zero projection of the orbital 
angular momentum (OAM) onto the propagation direction~%
\cite{MolinaTerriza_NP3_305:2007, Yao_OP3_161:2011, Torres:2011, AML:2013}.
These twisted (or vortex) beams possess helical phase wavefront and 
non-homogeneous intensity profile.
Due to these distinguishing features the twisted photons have found 
extensive applications, e.g. in optical~%
\cite{Bozinovic_S340_1545:2013} and free-space~%
\cite{Gibson_OE12_5448:2004, Wang_NP6_488:2012, Su_OE20_9396:2012} 
communications, metrology~\cite{Ambrosio_NC4_2432:2013}, and biophysics~%
\cite{Grier_N424_810:2003}.
Many of these applications require a detailed description of the 
fundamental atomic processes.
%
%
\\ \indent
%
%
During recent years numerous theoretical studies have been conducted to
investigate the effects of the twisted light beams in absorption~%
\cite{Schmiegelow_EPJD66_157:2012, Afanasev_PRA88_033841:2013, 
Scholz-Marggraf_PRA90_013425:2014, Surzhykov_PRA91_013403:2015, 
Schmiegelow_NC7_12998:2016, Muller_PRA94_041402:2016, 
Seipt_PRA94_053420:2016, Surzhykov_PRA94_033420:2016, 
Kaneyasu_PRA95_023413:2017} and scattering~\cite{Stock_PRA92_013401:2015, 
Zhang_PRL117_113904:2016} processes.
Much less attention has been paid to the question of whether emitted 
light is twisted or not.
The ``twistedness'' of the post-interaction photons has been estimated 
mainly in the processes being dedicated to their production.
As an example, the OAM of the emitted light was evaluated in the 
the Compton scattering~\cite{Jentschura:2011, Ivanov_PRD83_093001:2011, 
Ivanov_PRA84_033804:2011} and in the process of the high harmonic 
generation~\cite{Hernandez_PRL111_083602:2013, Rego_PRL117_163202:2016, 
Kong_NC8_14970:2017, Gauthier_NC8_14971:2017}.
The methods of these studies, however, are strongly related to the features 
of particular processes and cannot be extended to other situations.
To the best of our knowledge, no effort has been done to provide a 
theoretical approach which would allow to analyze the angular momentum 
properties of outgoing photons for arbitrary reaction.
%
%
\\ \indent
%
%
In this contribution we describe a simple theoretical method 
for the analysis of the angular momentum properties of the photons
emitted in fundamental atomic processes.
This method is based on the calculation of the average value of the
total angular momentum (TAM) projection operator of the outgoing photons.
The averaged value can be naturally calculated within the 
framework of the density matrix formalism.
This method allows one to find out whether the emitted photons are twisted 
or not for arbitrary reaction.
%
%
\\ \indent
%
%
We apply our method to analyze the angular momentum properties 
of photon beams for several cases.
First, the ``twistedness'' of the plane-wave, spherical-wave, and Bessel 
radiation has been re-explored.
As the second example, we analyze the angular momentum properties of light 
emitted due to the radiative recombination (RR) of electrons with bare nuclei.
We show that the RR photons, emitted along the electron beam direction, 
do carry a non-zero and well-defined projection of angular momentum.
%
%
\\ \indent
%
%
Relativistic units ($m_e = \hbar = c = 1$) and the Heaviside charge unit 
($e^2 = 4\pi\alpha$) are used in the paper.
%
%
%
%
%
%
%
\section{BASIC FORMALISM}
The main goal of the present paper is to formulate a theoretic method 
which will allow one to determine whether the photons emitted in basic 
atomic processes are twisted or not. 
For this purpose we start with the mathematical definition of the 
twisted light.
Here and throughout we restrict ourselves to the case of the Bessel 
twisted photons.
\subsection{Twisted photons}
\label{subs:tw_ph}
%
%
Let us consider the brief theoretical description of the Bessel-wave 
twisted photons.
These waves are the solutions of the free-wave equation in an empty 
space with the well-defined energy $\omega$, the helicity $\lambda$, 
and the projections of the momentum $k_z$ and total angular momentum 
(TAM) $m_\gamma$ onto the propagation direction.
This direction is chosen along the $z$ axis.
Additionally, the absolute value of the transverse momentum 
$\varkappa_\gamma = \left(\omega^2 - k_z^2\right)^{1/2}$ is well defined. 
Such a twisted photon state $\left\vert \varkappa_\gamma m_\gamma k_z
\lambda \right\rangle$
is described by the vector potential~\cite{Jentschura:2011, 
Ivanov_PRA84_033804:2011, Matula_JPB46_205002:2013}
\begin{eqnarray}
{\bf A}^{\rm (tw)}_{\varkappa_\gamma m_\gamma k_z\lambda}(\br) 
& = &
i^{\lambda - m_\gamma}
\int 
\frac{e^{im_\gamma\varphi_k} }{2\pi k_\perp} 
\delta(k_\parallel - k_z)
\delta(k_\perp - \varkappa_\gamma)
{\bf A}^{\rm (pl)}_{\bk\lambda}(\br)
d\bk,
\label{eq:vec_pot_tw}
\end{eqnarray}
where $k_\parallel$ and $k_\perp$ are the longitudinal and transversal
components of momentum $\bk$, respectively, and 
${\bf A}^{\rm (pl)}_{\bk\lambda}$ is the vector potential of the 
plane-wave photon 
\begin{equation}
{\bf A}^{\rm (pl)}_{\bk\lambda}(\br) 
= \frac{\bepsilon_\lambda(\bk) e^{i\bk\cdot\br}}{\sqrt{2\omega(2\pi)^3}}.
\label{eq:vec_pot_pw}
\end{equation}
Eq.~\eqref{eq:vec_pot_tw} implies that the Bessel light can be ``seen'' 
as a coherent superposition of the plane-wave photons with the linear 
momenta $\bk$ laying on the surface of a cone with the opening 
angle $\theta_\gamma = \arctan\left(\varkappa_\gamma / k_z\right)$.
\\
\indent
In the literature one may find many definitions of the twisted light.
Here we will term photons as twisted, in the sense of pure Bessel beams,
if they possess a well-defined TAM projection and a well-defined opening 
angle differing from $0^\circ$.
Therefore, in order to determine the OAM properties of the photon one 
needs to calculate its TAM projection and the opening angle.
Instead, of the evaluation of the opening angle one can calculate its'
sine or cosine.
For the monochromatic photon beam, the evaluation of the opening angle 
cosine simplifies to the calculation of the longitudinal momentum.
In the framework of the present investigation we restrict our 
consideration to this type of beams.
%
%
\subsection{Evaluation of TAM projection and opening angle of light}
\label{subs:anal}
%
%
As described above, the twisted light is characterized by the TAM 
projection and by the opening angle.
Below we consider a method of the evaluation of the mean values of 
these two quantities.
In the previous section it was assumed that the propagation direction of 
the twisted light coincides with the $z$-axis. 
But this is not always the case for the atomic processes. 
We analyze, therefore, the TAM projection onto the propagation direction
of the photons emitted in some arbitrary $\hat{\bn}_0$ direction.
The mean values of the TAM projection operator and the opening angle are 
conveniently evaluated within the framework of the density matrix formalism.
In this approach, the average value of the projection of the TAM 
operator $\bJ$ onto some arbitrary $\hat{\bn}_0$ axis, defining the 
propagation direction of the emitted photons, is given by
\begin{equation}
\left\langle \bJ\cdot \hat{\bn}_0 \right\rangle 
= \frac{Tr\left[\rho^{(\rm ph)} \rho_{\hat{\bn}_0}^{(\rm det)} \tam\right]}
{Tr\left[\rho^{(\rm ph)} \rho^{(\rm det)}_{\hat{\bn}_0} \right]},
\label{eq:tam_def}
\end{equation}
where $\rho^{(\rm ph)}$ is the density operator of the photon and the 
operator $\rho_{\hat{\bn}_0}^{(\rm det)}$ describes the detector.
The form of the detector operator depends on a particular experiment. 
In our study we consider so large detector that it can be approximated by 
a plane-wave detector located perpendicular to the $\hat{\bn}_0$ direction.
%
%
\\ \indent
%
%
The right-hand side of Eq.~\eqref{eq:tam_def} is written in the operator 
form.
For practical applications it is more convenient to re-write this 
expression in the matrix form, which requires choosing the basis 
representation of photons states.
Here we use the helicity basis of plane-wave solutions, 
$\left\vert\bk\lambda\right\rangle$, where $\bk$ is the wave vector and 
$\lambda$ is the helicity, in which the expression~\eqref{eq:tam_def} is 
given by
\begin{equation}
\left\langle \bJ\cdot \hat{\bn}_0 \right\rangle 
= 
\frac{
\sum_{\lambda\lambda'\lambda''}
\int d\bk d\bk' d\bk'' \left(8\omega\omega'\omega''\right)
\left\langle \bk\lambda \left\vert \rho^{(\rm ph)} \right\vert \bk'\lambda' \right\rangle
\left\langle \bk'\lambda' \left\vert \rho^{(\rm det)}_{\hat{\bn}_0} \right\vert \bk''\lambda'' \right\rangle
\left\langle \bk''\lambda'' \left\vert \tam \right\vert \bk\lambda \right\rangle
}{
\sum_{\lambda\lambda'} 
\int d\bk d\bk' \left(4\omega\omega'\right)
\left\langle \bk\lambda \left\vert \rho^{(\rm ph)} \right\vert \bk'\lambda' \right\rangle
\left\langle \bk'\lambda' \left\vert \rho^{(\rm det)}_{\hat{\bn}_0} \right\vert \bk\lambda \right\rangle}.
\label{eq:tam_def_pw}
\end{equation}
The states $\left\vert\bk\lambda\right\rangle$ are described by the 
vector potential~\eqref{eq:vec_pot_pw} and satisfy the following 
completeness condition
\begin{equation}
\sum_\lambda\int d\bk \left(2\omega\right)
\left\vert\bk\lambda\right\rangle\left\langle\bk\lambda\right\vert = I,
\end{equation}
with $I$ being the unity operator.
In the helicity basis of plane-wave solutions the matrix element of the 
detector operator expresses as~\cite{Balashov}
\begin{equation}
\left\langle \bk'\lambda' \left\vert \rho^{(\rm det)}_{\hat{\bn}_0} \right\vert \bk\lambda \right\rangle
=
\frac{1}{2\omega}
\delta\left(\bk - \bk'\right)
\theta\left(\bk\cdot\hat{\bn}_0\right)
\delta_{\lambda\lambda'},
\label{eq:det_oper}
\end{equation}
where $\theta(x)$ is the Heaviside function.
Substituting Eq.~\eqref{eq:det_oper} into Eq.~\eqref{eq:tam_def_pw} one
obtains the following expression for the average value of the TAM 
projection operator
\begin{equation}
\left\langle \bJ\cdot \hat{\bn}_0 \right\rangle 
= 
\frac{
\sum_{\lambda\lambda'}
\int d\bk d\bk' \left(4\omega\omega'\right)
\left\langle \bk\lambda \left\vert \rho^{(\rm ph)} \right\vert \bk'\lambda' \right\rangle
\left\langle \bk'\lambda' \left\vert \tam \right\vert \bk\lambda \right\rangle
\theta\left(\bk\cdot\hat{\bn}_0\right)
}{
\sum_\lambda
\int d\bk \left(2\omega\right)
\left\langle \bk\lambda \left\vert \rho^{(\rm ph)} \right\vert \bk\lambda \right\rangle
\theta\left(\bk\cdot\hat{\bn}_0\right)
}.
\label{eq:tam_def_det}
\end{equation}
The explicit form of the photon density matrix $\left\langle \bk\lambda 
\left\vert \rho^{(\rm ph)} \right\vert \bk'\lambda' \right\rangle$ depends 
on the particular ``scenario'' under investigation.
In the present paper we consider the cases of plane-wave, spherical-wave, 
and Bessel radiation as well as of RR photons.
%
%
\\
\indent
%
%
While the evaluation of the photon density matrix requires the knowledge 
about a process under consideration, the matrix element of the operator $\tam$, 
which also enters into Eq.~\eqref{eq:tam_def_pw} is independent on the
particular ``scenario''.
It is conveniently calculated in the momentum representation for the TAM 
operator~\cite{Akhiezer}
\begin{equation}
{\bf J}_p = -i\left[\bp \times \nabla_p\right] + {\bf S},
\label{eq:tam_mom}
\end{equation}
where ${\bf S}$ is the spin-1 operator. 
Apart from the operator $\bJ$, the vector potential of the plane-wave 
photon has also to be written in the momentum representation:
\begin{equation}
{\bf f}^{(\rm pl)}_{\bk\lambda}(\bp) = \frac{\bepsilon_\lambda(\bk)}{\sqrt{2\omega}}\delta(\bp - \bk),
\label{eq:vec_pot_pw_mom}
\end{equation}
which is related to the vector potential in the coordinate representation~%
\eqref{eq:vec_pot_pw} by the following simple relation
\begin{equation}
{\bf A}^{(\rm pl)}_{\bk\lambda}(\br) 
= \frac{1}{\sqrt{(2\pi)^3}}\int d\bp\
{\bf f}^{(\rm pl)}_{\bk\lambda}(\bp) e^{i\bp\cdot\br}.
\end{equation}
Utilizing Eqs.~\eqref{eq:tam_mom} and~\eqref{eq:vec_pot_pw_mom}, one can 
derive the explicit expression for the matrix element of the TAM 
projection operator
\begin{eqnarray}
\nonumber
\left\langle \bk'\lambda' \left\vert 
\tam
\right\vert \bk\lambda \right\rangle
& = & 
\int d\bp 
{\bf f}^{(\rm pl)\dagger}_{\bk'\lambda'}(\bp)
\left( {\bf J}_p\cdot\hat{\bn}_0 \right)
{\bf f}^{(\rm pl)}_{\bk\lambda}(\bp)
=
\frac{1}{4\pi k^2}\delta(k - k')\frac{\delta_{\lambda\lambda'}}{2\omega}
\sum_{\mu} \left(\hat{\bn}_0\right)^{\mu} 
\\ && \times
\sum_{jm_jm_j'} (2j + 1)
\sqrt{j(j + 1)} C_{jm_j\, 1\mu}^{jm_j'} 
D_{m_j\lambda}^{j}(\varphi_k, \theta_k, 0)
D_{m'_j\lambda}^{j*}(\varphi_k', \theta_k', 0).
\label{eq:tam_pw_me}
\end{eqnarray}
Here $(\hat{\bn}_0)^{\mu}$ are the contravariant vector components, 
$C_{j_1m_1\, j_2m_2}^{JM}$ is the Clebsch-Gordan coefficient, $D_{MM'}^J$ 
is the Wigner matrix~\cite{Rose_ETAM, Varshalovich}, $(k,\theta_k,\varphi_k)$ 
are the spherical coordinates of $\bk$, and $(k',\theta_k',\varphi_k')$ 
are those of $\bk'$.
\\
\indent
Substituting the explicit form of $\left\langle \bk\lambda \left\vert 
\rho^{(\rm ph)} \right\vert \bk'\lambda' \right\rangle$ and 
Eq.~\eqref{eq:tam_pw_me} into Eq.~\eqref{eq:tam_def_pw}, one can evaluate
the average value of the TAM projection.
But the mean value of the TAM projection operator can not solely describe 
the ``twistedness'' of light.
Indeed, in accordance with the definition (see Section~\ref{subs:tw_ph}),
the light is called twisted if its TAM projection onto the propagation 
direction is well-defined. 
Therefore, one needs to know not only the mean value but also the 
dispersion of TAM
\begin{equation}
\Delta_J = 
\sqrt{
\left\langle \tam^2 \right\rangle - \left\langle \bJ\cdot \hat{\bn}_0 \right\rangle^2
}.
\label{eq:tam_disp_def}
\end{equation}
As is seen from this expression, the evaluation of $\Delta_J$
requires the knowledge of not only $\left\langle \bJ\cdot \hat{\bn}_0 
\right\rangle$ given by Eq.~\eqref{eq:tam_pw_me} but also of 
$\left\langle \tam^2 \right\rangle$.
By using Eqs.~\eqref{eq:tam_mom} and~\eqref{eq:vec_pot_pw_mom} and 
performing some tedious but straightforward calculations, one obtains 
the explicit expression for the matrix element of the $\tam^2$ operator.
For the sake of brevity we will omit details of these calculations 
here and just present the final result
\begin{eqnarray}
\nonumber
\left\langle \bk'\lambda' \left\vert 
\tam^2
\right\vert \bk\lambda \right\rangle
& = &
\frac{1}{\sqrt{4\pi} k^2}\delta(k - k')\frac{\delta_{\lambda\lambda'}}{2\omega}
\sum_{J_n M_n} C_{10\, 10}^{J_n0} 
Y^*_{J_nM_n}(\hat{\bn}_0)
\sum_{jm_jm_j'} j(j + 1)(2j + 1)^{3/2}
\\ && \times 
C_{jm_j\, J_nM_n}^{jm_j'} 
\begin{Bmatrix}
1 & 1 & J_n
\\
j & j & j
\end{Bmatrix}
D_{m_j\lambda}^{j}(\varphi_k, \theta_k, 0)
D_{m'_j\lambda}^{j*}(\varphi_k', \theta_k', 0).
\end{eqnarray}
Here $\{ \cdots\}$ denotes the Wigner $6j$ symbol~\cite{Varshalovich} and 
$Y_{lm}(\theta,\varphi)$ is the spherical harmonic.
%
%
\\
\indent
%
%
Up to now we have discussed the evaluation of the mean value and the 
dispersion of the TAM projection of light.
As was already mentioned, in order to determine whether the emitted 
photon is twisted or not one needs also to evaluate the opening angle 
$\theta_\gamma$ or its cosine.
In the case of the monochromatic photon beam 
\begin{equation}
\cos\theta_\gamma = \frac{1}{\omega} 
\left\langle {\bf p} \cdot \hat{\bn}_0\right\rangle,
\label{eq:cos_theta}
\end{equation}
where ${\bf p}$ is the momentum operator.
Re-writing the expression~\eqref{eq:cos_theta} in the form similar to 
Eq.~\eqref{eq:tam_def_det} and utilizing the explicit form of the matrix 
elements:
\begin{eqnarray}
\left\langle \bk'\lambda' \left\vert 
\mom
\right\vert \bk\lambda \right\rangle & = &
\frac{\delta_{\lambda\lambda'}}{2\omega} 
\delta(\bk - \bk')
\left(\bk \cdot \hat{\bn}_0\right),
\\
\left\langle \bk'\lambda' \left\vert 
\mom^2
\right\vert \bk\lambda \right\rangle & = &
\frac{\delta_{\lambda\lambda'}}{2\omega} 
\delta(\bk - \bk')
\left(\bk \cdot \hat{\bn}_0\right)^2.
\end{eqnarray}
one can evaluate the mean value and the opening angle cosine 
$\cos\theta_\gamma$.
The dispersion $\Delta_p$ is defined analogously to the dispersion 
$\Delta_J$.
%
%
%
%
%
%
%
%
%
%
\section{RESULTS AND DISCUSSIONS}
%
%
\subsection{TAM and its dispersion for plane-wave, spherical-wave, and 
Bessel photons}
%
%
In order to demonstrate the method which is described above let us 
evaluate the TAM projection, momentum projection, which is directly 
related to the opening angle cosine~\eqref{eq:cos_theta}, and their 
dispersions for the plane-wave, spherical-wave, and twisted photons.
%
%
\subsubsection{Plane-wave photons}
%
%
As was discussed in Section~\ref{subs:anal}, in order to find the 
average value of the TAM and its dispersion it is sufficient to 
calculate the trace of the density matrix with TAM and squared TAM 
projection operators.
The density operator for the plane-wave photon with the momentum $\bk$ and 
polarization $\bepsilon_\lambda$ is given by
\begin{equation}
\rho_{\bk\lambda}^{\rm (pl)}  = 
\left\vert \bk\lambda\right\rangle
\left\langle \bk\lambda \right\vert.
\label{eq:dens_mtrx_pw}
\end{equation}
In the present study we restrict ourselves to the case of TAM projection 
onto the photon propagation direction, i.e. $\hat{\bn}_0 = 
\hat{\bk}\equiv\bk/|\bk|$.
For this case one obtains:
\begin{eqnarray}
\left\langle\bJ\cdot \hat{\bk}\right\rangle_{\rm pl} = \lambda,
\quad
\left\langle\left(\bJ\cdot \hat{\bk}\right)^2\right\rangle_{\rm pl} = 1,
\quad
\Delta^{(\rm pl)}_J = 0,
\label{eq:pw_tam}
\\
\left\langle\bp\cdot \hat{\bk}\right\rangle_{\rm pl} = \omega,
\quad
\left\langle\left(\bp\cdot \hat{\bk}\right)^2\right\rangle_{\rm pl} = \omega^2,
\quad
\Delta^{(\rm pl)}_p = 0.
\label{eq:pw_mom}
\end{eqnarray}
The formulas~\eqref{eq:pw_tam} represent the well-known fact that the 
TAM projection of the plane-wave photon on its propagation direction is 
given by the helicity $\lambda$.
The expressions~\eqref{eq:pw_mom} indicate that the plane-wave photon is 
the eigenfunction of the $\bp$ operator.
%
%
\subsubsection{Spherical-wave photons}
%
%
The density operator for the spherical-wave photon with energy $\omega$, 
TAM $j$, and TAM projection onto the $z$ axis $m_\gamma$ is given by
\begin{equation}
\rho_{\omega j m_\gamma \pi}^{\rm (sph)}  = 
\left\vert \omega j m_\gamma \pi\right\rangle
\left\langle \omega j m_\gamma \pi \right\vert, 
\label{eq:dens_mtrx_sw}
\end{equation}
with $\pi = 0$ for the magnetic and $\pi = 1$ for the electric photon.
The explicit form of the vector potential of the spherical photon in the 
momentum space expresses as follows~\cite{Berestetsky}
\begin{equation}
{\bf f}^{\rm (sph)}_{\omega j m_\gamma \pi}(\bp) = 
\frac{4\pi^2}{\omega^{3/2}} \delta(|\bp| - \omega) 
{\bf Y}^{(\pi)}_{jm_\gamma}(\hat{\bp}),
\label{eq:vec_pot_sw_mom}
\end{equation}
where ${\bf Y}^{(\pi)}_{jm_\gamma}$ is the spherical harmonic vectors~%
\cite{Varshalovich}.
Utilizing the formalism described in Section~\ref{subs:anal} and Eqs.
\eqref{eq:dens_mtrx_sw} and~\eqref{eq:vec_pot_sw_mom}, one can 
calculate the average value of the projection of the TAM operator onto 
some arbitrary $\hat{\bn}_0$ axis.
Here we focus on the situation when $\hat{\bn}_0$ coincides with the 
quantization $z$ axis, i.e. $\hat{\bn}_0 = \hat{\bf e}_z$ with 
$\hat{\bf e}_z$ being the unit vector directed along the $z$ axis.
In this case:
\begin{equation}
\left\langle J_z\right\rangle_{\rm sph} = m_\gamma,
\quad
\left\langle J_z^2\right\rangle_{\rm sph} = m^2_\gamma,
\quad
\Delta^{(\rm sph)}_J = 0.
\end{equation}
From these equations one can see that the spherical-wave 
photon is the eigenfunction of the $J_z$ operator with the eigenvalue 
$m_\gamma$.
It is worth mentioning that the average value of the opening angle 
cosine and its' dispersion are both depend on the $j$ and $m_\gamma$.
And since these dependencies cannot be expressed by a compact formula we 
omit them in the sake of brevity.
%
%
\subsubsection{Twisted photons}
%
%
Let us now consider the case of the Bessel-wave twisted photon 
propagating along the $z$ axis.
The corresponding density operator is given by
\begin{equation}
\rho_{\varkappa_\gamma m_\gamma k_z \lambda}^{\rm (tw)}  = 
\left\vert \varkappa_\gamma m_\gamma k_z \lambda \right\rangle
\left\langle \varkappa_\gamma m_\gamma p_z \lambda \right\vert,
\label{eq:dens_tw}
\end{equation}
where $\varkappa_\gamma$ and $k_z$ are the transversal and longitudinal 
momenta, $\lambda$ is the helicity, and $m_\gamma$ is the TAM projection
onto the propagation direction.
As in the case of the plane- and spherical-wave photons, we restrict 
ourselves to evaluation of TAM projection and its dispersion for the 
particular direction of $\hat{\bn}_0$.
Namely, we study the situation when $\hat{\bn}_0$ is directed along the
propagation direction, i.e. $\hat{\bn}_0 = \hat{\bf e}_z$.
In this case one obtains
\begin{eqnarray}
\left\langle J_z\right\rangle_{\rm tw} = m_\gamma,
\quad
\left\langle J_z^2\right\rangle_{\rm tw} = m^2_\gamma,
\quad
\Delta^{(\rm tw)}_J = 0,
\\
\left\langle p_z \right\rangle_{\rm tw} = k_z,
\quad
\left\langle p_z^2 \right\rangle_{\rm tw} = k_z^2,
\quad
\Delta^{(\rm tw)}_p = 0.
\label{eq:tw_mom}
\end{eqnarray}
As is expected from the form of the density operator~\eqref{eq:dens_tw}, 
the mean value of the TAM projection onto the propagation direction of 
the twisted photon equals $m_\gamma$ with the zero dispersion.
The formulas~\eqref{eq:tw_mom} denote that the Bessel-wave twisted photon
is the eigenfunction of the $p_z$ operator.
%
%
\subsection{Radiative recombination of electrons with bare nuclei}
%
%
Until now we have applied our approach to study the ``twistedness'' of 
light to the textbook examples of the plane-wave, spherical-wave, and 
Bessel-wave twisted radiation.
Let us now turn to the analysis of the photons emitted in one of the 
fundamental processes of light-matter interaction, namely the radiative 
recombination (RR) of electrons with bare nuclei.
Despite a large number of studies devoted to this process (for a review 
see Ref.~\cite{Eichler_PR439_1:2007}) no attention has been paid so far 
to the angular momentum properties of the RR photons.
Below we analyze these properties for two different scenarios.
In the first one we assume that the incident electrons are prepared
in the plane-wave state, while in the second one in the twisted one.
For the second scenario the collisions with a single ion and a macroscopic 
target are considered.
%
%
\subsubsection{Recombination of plane-wave electrons}
\begin{figure}[h!]
\centering
\includegraphics{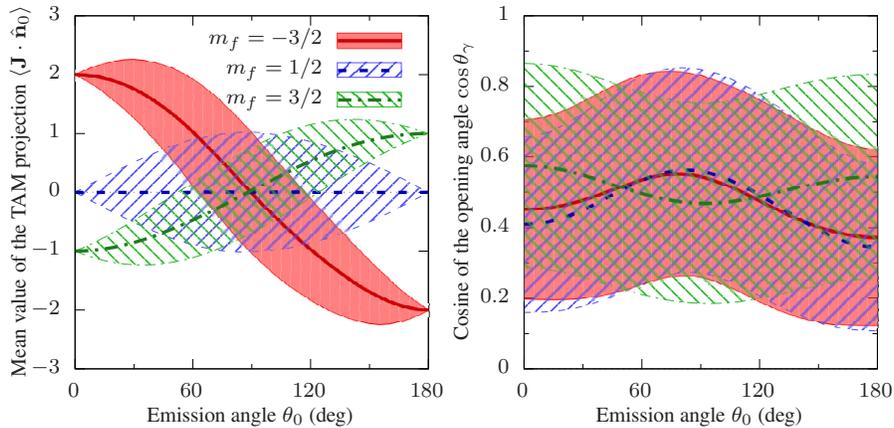} 
\caption{
The mean value of the operator of the TAM projection on the direction of
photon emission $\hat{\bn}_0 = \left(\sin\theta_0, 0, \cos\theta_0\right)$
(left panel) and the cosine of the opening angle (right panel).
The recombination of the 2 keV plane-wave electron with $\mu = 1/2$ into 
the $2p_{3/2}(m_f)$ state of the H-like Ar ($Z = 18$) ion is considered. 
The shadowed areas designate the dispersions.
}
\label{ris:plane_wave}
\end{figure}

%
%
Let us start from the simplest case, the recombination of a plane-wave 
electron with a bare nucleus.
The density matrix of the photons emitted in course of the RR of the 
asymptotically plane-wave electron with the momentum $\bp$ and the 
helicity $\mu$ into the final bound $f$ state with the TAM projection 
$m_f$ has the following form
\begin{equation}
\left\langle \bk\lambda 
\left\vert \rho^{(\rm pl)}_{\bp\mu;fm_f} \right\vert 
\bk'\lambda' \right\rangle
=
\tau^{(\rm pl)}_{\bp\mu;fm_f,\bk\lambda}
\tau^{(\rm pl)*}_{\bp\mu;fm_f,\bk'\lambda'}
\label{eq:density_rr_pl},
\end{equation}
with the amplitude
\begin{equation}
\tau^{(\rm pl)}_{\bp\mu;fm_f,\bk\lambda} = \int d\br 
\Psi_{fm_f}^\dagger(\br)
R^\dagger_{\bk\lambda}(\br)
\Psi_{\bp\mu}^{(+)}(\br).
\end{equation}
Here $\Psi_{fm_f}$ is the wave function of the electron in the final 
state, $R_{\bk\lambda}$ designates the transition operator which has 
the following form in the Coulomb gauge
\begin{equation}
R_{\bk\lambda}(\br) = -\sqrt{\frac{\alpha}{\omega(2\pi)^2}}
\balpha\cdot\bepsilon_\lambda e^{i\bk\cdot\br},
\end{equation}
with $\balpha$ being the vector of Dirac matrices, and 
$\Psi_{\bp\mu}^{(+)}$ is the wave function of the electron in the initial 
state given by~\cite{Eichler, Rose_RET, Pratt}
\begin{equation}
\Psi_{\bp\mu}^{(+)}(\br) = \frac{1}{\sqrt{4\pi\varepsilon p}} \sum_{\kappa m_j}
C_{l0\,1/2\mu}^{j\mu} i^l \sqrt{2l+1} e^{i\delta_\kappa}
D_{m_j\mu}^j\left(\varphi_p,\theta_p,0\right)
\Psi_{\varepsilon\kappa m_j}(\br).
\label{eq:wf_pw}
\end{equation}
Here $\kappa = (-1)^{l+j+1/2}(j + 1/2)$ is the Dirac quantum number with 
$j$ and $l$ being the TAM and OAM, respectively, and $\delta_\kappa$ is 
the phase shift corresponding to the potential of the extended nucleus.
%
%
\\ 
\indent
%
%
Above we have presented the density matrix~\eqref{eq:density_rr_pl} of 
the photon emitted in course of the RR of a plane-wave electron with a 
bare nucleus.
Now we turn to the evaluation of ``twistedness'' of this radiation.
Let us fix the $z$ axis along the propagation direction of the 
incoming electron.
For such choice of the coordinate system the TAM projection onto the $z$ 
axis, i.e. $\hat{\bn}_0 = \hat{\bf e}_z$, and its dispersion equal, 
respectively,
\begin{equation}
\left\langle J_z \right\rangle_{\rm pl} = \mu - m_f,
\quad
\Delta_J = 0.
\label{eq:jz_pw}
\end{equation}
This equation indicates that the photons being emitted in course 
of the RR of the polarized plane-wave electron and propagating in the 
forward direction do possess the well-defined projection of TAM onto 
their propagation direction.
This means that the RR photons can carry the nonzero projection 
of the OAM onto the propagation direction which is determined solely by 
the helicity $\mu$ of the incident electron and by the magnetic quantum 
number of the residual ion $m_f$.
%
%
\\ 
\indent
%
%
Above we analyzed the angular momentum properties of the photons
emitted along the $z$ axis.
Let us remind here that the $z$ axis is fixed along the propagation 
direction of the incoming electron.
Now let us consider the angular momentum properties of the RR photons 
emitted into some arbitrary direction $\hat{\bn}_0 \neq \hat{\bf e}_z$.
This case is represented in Fig.~\ref{ris:plane_wave} for the RR with 
the bare argon nuclei.
On the left panel of this figure it is seen that for the forward and 
backward emission angles the TAM projection takes the well-defined 
values.
This fact is predicted by the relation~\eqref{eq:jz_pw}.
From the right panel of Fig.~\ref{ris:plane_wave} one can conclude that 
for all propagation directions the emitted photons do not have the 
well-defined opening angle $\theta_\gamma$ and consequently transversal 
momentum.
This can be explained as follows.
In the external field of the nucleus the momentum does not conserve and,
as a result, the distribution of the momentum occurs.
In accordance with definition given in Section~\ref{subs:tw_ph},
the RR photons can not be regarded as twisted.
But, these photons can neither be regarded as the plane or the spherical
wave since the cosine of the opening angle always differs from 1 
and 0, respectively (see the right panel of Fig.~\ref{ris:plane_wave}).
Therefore, one can say that the RR photons emitted in the forward or 
backward directions are, in some sense, twisted.
%
%
\subsubsection{Recombination of twisted electrons}
\begin{figure}[h!]
\centering
\includegraphics{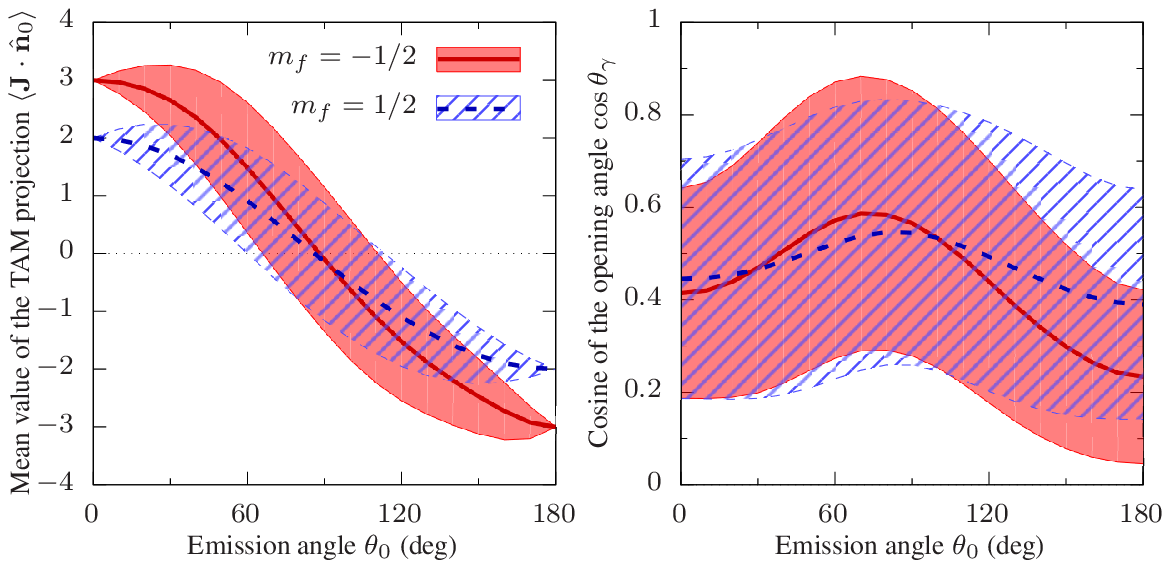} 
\caption{
The mean value of the operator of the TAM projection on the direction of
photon emission $\hat{\bn}_0 = \left(\sin\theta_0, 0, \cos\theta_0\right)$
(left panel) and the cosine of the opening angle (right panel).
The recombination of the 2 keV twisted electron with $m_e = 5/2$, $\mu = 1/2$, 
and $\theta_p = 30^\circ$ into the ground $1s(m_f)$ state of the H-like 
Ar ($Z = 18$) ion is considered. 
It is assumed that the ion is placed on the $z$ axis ($\bb = 0$).
The shadowed areas designate the dispersions of the average values.
}
\label{ris:sngl_b0}
\end{figure}

\begin{figure}[h!]
\centering
\includegraphics{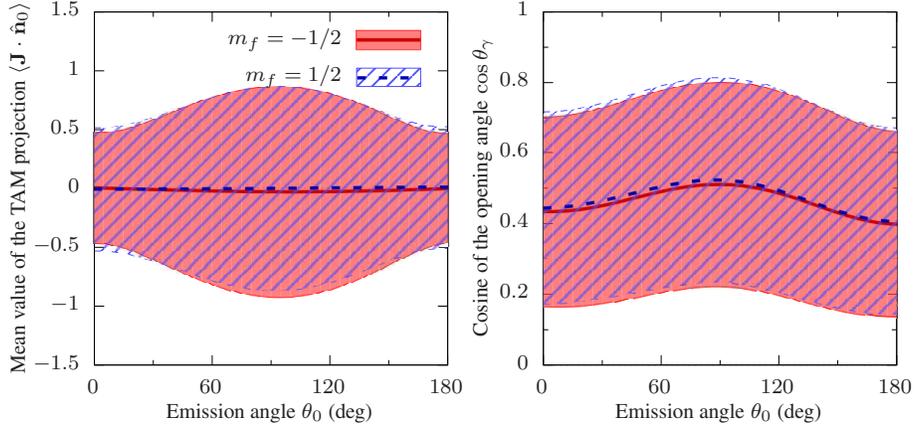} 
\caption{
The mean value of the operator of the TAM projection on the direction of
photon emission $\hat{\bn}_0 = \left(\sin\theta_0, 0, \cos\theta_0\right)$
(left panel) and the cosine of the opening angle (right panel).
The recombination of the 2 keV twisted electron with $m = 5/2$, $\mu = 1/2$, 
and $\theta_p = 30^\circ$ into the ground $1s(m_f)$ state of the H-like 
Ar ($Z = 18$) ion is considered. 
It is assumed that the ion is shifted along the $x$ axis from the 
electron propagation direction on 1 nm.
The shadowed areas designate the dispersions of the average values.
}
\label{ris:sngl_b1}
\end{figure}

%
%
Up to now we have discussed recombination of the plane-wave electrons. 
Nowadays, one can also use the twisted electrons instead of the 
conventional ones.
These electrons possessing a non-zero projection of OAM onto their 
propagation direction can be readily produced with present experimental 
techniques~\cite{Verbeeck_N467_301:2010, Uchida_N464_737:2010, 
McMorran_S331_192:2011, Grillo_PRL114_034801:2015, Mafakheri_MM21_667:2015}.
It is of interest, therefore, to investigate the possibility of the TAM transfer 
from the twisted electron beam to the RR photons.
Previously, the recombination of the vortex electrons was studied in 
Refs.~\cite{Matula_NJP16_053024:2014, Zaytsev_PRA95_012702:2017}.
In both works, however, no attention has been paid to the question 
whether the emitted radiation is twisted or not.
Below we evaluate the TAM projection of the radiative photons and 
thereby fill in the gap.
\\
\indent
Let us start with the brief recall of the main properties of the free 
twisted electrons which we take here in the form of the Bessel waves.
As the vortex photons, these electrons are characterized by the 
following set of quantum numbers.
The energy $\varepsilon$, the helicity $\mu$, and the projections of the 
linear $p_z$ and total angular $m_e$ momenta onto the propagation direction 
which is chosen as the $z$ axis.
Twisted electrons possess a well-defined transversal 
momentum $\varkappa_e = \sqrt{\varepsilon^2 - 1 -p_z^2}$ and the so-called
opening angle $\theta_e = \arctan\left(\varkappa_e / p_z\right)$.
The wave function of the state with the quantum numbers listed above 
have an inhomogeneous probability distribution and an inhomogeneous 
probability current density (see, e.g., Ref.~\cite{Zaytsev_PRA95_012702:2017}).
Due to this feature, the relative position of the target with respect to
the scattering electron and the type of the target become important.
Here we consider two types of targets, viz. a single ion and an infinitely 
extended (macroscopic) target.
%
%
\paragraph{Single ion}\mbox{}\\
%
%
Having discussed the basic properties of the twisted electrons, we are 
ready to study their interaction with ionic targets. 
First, we will consider the case of electron recombination with a single 
bare nucleus placed at some well-defined position inside the vortex
electron beam.
The density matrix for such a scenario can be written as
\begin{equation}
\left\langle \bk\lambda 
\left\vert \rho^{(\rm tw, sngl)}_{\varkappa_e m_e p_z \mu; fm_f}(\bb) \right\vert 
\bk'\lambda' \right\rangle
=
\tau^{(\rm tw)}_{\varkappa_e m_e p_z \mu;fm_f,\bk\lambda}(\bb)
\tau^{(\rm tw)*}_{\varkappa_e m_e p_z \mu;fm_f,\bk'\lambda'}(\bb).
\label{eq:density_rr}
\end{equation}
The amplitude of this process can be constructed from the plane-wave 
electron RR amplitude as follows~\cite{Zaytsev_PRA95_012702:2017}
\begin{equation}
\tau^{(\rm tw)}_{\varkappa_e m_e p_z \mu;fm_f,\bk\lambda}(\bb)
=
i^{\mu - m_e}\int\frac{e^{im_e\varphi_p}}{2\pi p_\perp}
\delta(p_\parallel - p_z) \delta(p_\perp - \varkappa_e) 
e^{i\bp\cdot\bb} \tau^{(\rm pl)}_{\bp\mu;fm_f,\bk\lambda} d\bp,
\label{eq:amplitude_tw}
\end{equation}
where $\bb = \left(b\cos\varphi_b,b\sin\varphi_b,0\right)$ is the impact 
parameter, i.e. the distance from the target ion to the central ($z$) 
axis of the incident vortex electron beam.
Substituting the density matrix~\eqref{eq:density_rr} into 
Eq.~\eqref{eq:tam_def} one can evaluate the mean value of the TAM 
operator projection onto the direction of the photon emission $\hat{\bn}_0$ 
and thereby find out whether the emitted photons are twisted or not.
%
%
\\ 
\indent
%
%
The most interesting situation occurs when the ion is placed in the 
center of the vortex electron beam ($\bb = 0$).
In this case, it can be analytically shown that for the forward photon 
emission, $\hat{\bn}_0 = \hat{\bf e}_z$,
\begin{equation}
\left\langle J_z\right\rangle^{(b = 0)}_{\rm tw}
= 
m_e - m_f,
\quad
\Delta^{\rm (b = 0)}_J = 0.
\label{eq:tam_tw_sngl}
\end{equation}
This means that the radiative photons propagating along the $z$ axis do 
carry the well-defined projection of the TAM onto their propagation 
direction.
And this projection is determined solely by the TAM projections of the 
initial $m_e$ and final $m_f$ electrons states.
This is explained as follows.
For $\bb = 0$ the entire system possesses the azimuthal symmetry with 
respect to the $z$ axis.
As a result, the angular momentum projection on this axis is conserved and 
the TAM projection of the twisted electron $m_e$ is equal to the sum of 
$m_f$ and TAM projection of the emitted photon.
This is expressed by Eq.~\eqref{eq:tam_tw_sngl}
%
%
\\
\indent
%
%
The angular momentum properties of the RR photons emitted into some 
arbitrary direction $\hat{\bn}_0 \neq \hat{\bf e}_z$ are presented in 
Fig.~\ref{ris:sngl_b0}.
From the left panel of this figure one can see that the photons emitted 
in the directions $\hat{\bn}_0 \neq \pm\hat{\bf e}_z$ do not possess a 
well-defined value of the TAM projection onto the propagation direction.
From the right panel of Fig.~\ref{ris:sngl_b0} it is seen that the 
emitted photons do not have a well-defined opening angle $\theta_\gamma$.
However, as in the case of the plane-wave electron recombination,
we can say that the RR photons emitted in the forward or backward 
directions are, in some sense, twisted.
%
%
\\ 
\indent
%
%
Up to now we discussed the case when the ion was placed on the electron 
vortex line, $\bb = 0$.
If the ion is displaced from this axis by the impact parameter $\bb$, the 
rotational symmetry is broken.
In this case, the projection of the TAM of the RR photon is not well-defined. 
It is clearly seen from Fig.~\ref{ris:sngl_b1}, where the results for the 
twisted electron RR with the bare argon nucleus being shifted from the $z$ 
axis are depicted.
It is also worth mentioning that the dependence on the TAM projection of 
the incident twisted electron $m_e$ is almost absent.
%
%
\paragraph{Macroscopic target}
\mbox{}\\
\indent
%
%
The single-ion target is interesting from theoretical viewpoint but it 
can not be realized in experiment.
We consider, therefore, a more realistic scenario in which the twisted 
electron beam collides with a macroscopic target, which we describe as 
an incoherent superposition of ions being homogeneously distributed.
The density matrix for this case is given by~\cite{Zaytsev_PRA95_012702:2017}
\begin{equation}
\left\langle \bk\lambda 
\left\vert \rho^{(\rm tw, mac)}_{\varkappa_e m_e p_z \mu; fm_f} \right\vert 
\bk'\lambda' \right\rangle
=
\int \frac{d\bb}{\pi R^2}
\left\langle \bk\lambda 
\left\vert \rho^{(\rm tw, sngl)}_{\varkappa_e m_e p_z \mu; fm_f}(\bb) \right\vert 
\bk'\lambda' \right\rangle,
\label{eq:density_macr}
\end{equation}
where $1 / (\pi R^2)$ is the cross section area with $R$ being the radius 
of the cylindrical box. 
In the case of the twisted electron RR with the macroscopic target the 
emitted photon, unfortunately, does have neither a well-defined TAM 
projection nor a well-defined opening angle.
%
%
%
%
%
%
%
%
\section{CONCLUSION}
%
%
In the present work, we described the simple theoretical method for the
evaluation of the ``twistedness'' of the photons emitted in basic atomic  
processes.
As the applications of the proposed method, we evaluated the 
TAM projection and its dispersion for the plane-wave, spherical-wave, 
and twisted photons.
We have also analyzed the ``twistedness'' of the photons emitted 
in the radiative recombination of electrons with the bare argon nuclei.
Two different situations have been considered.
In the first scenario it was assumed that the incident electron was prepared 
in the plane-wave state, while the twisted state was considered in the 
second one.
It was found that in the first scenario the RR photons emitted in the 
forward or backward directions have the well-defined TAM projection onto 
this direction.
For these photons the TAM projection is determined solely by the 
polarizations of the incident plane-wave electron and by the magnetic 
quantum number of the residual ion.
In the second scenario, the most interesting result has been obtained 
for the RR with the ion placed in the center of the electron beam.
In this case the recombination radiation propagating in the forward 
direction does possess a well-defined TAM projection onto this direction.
This result does not retain for the target ion being shifted from the 
propagation direction as well as for the recombination with the 
macroscopic target.
And, although, for the both scenarios the emitted photons do not have 
well-defined opening angles, we believe that the RR photons for the forward 
or backward emission directions are, in some sense, twisted.
%
%
\\ \indent
%
%
%
To summarize, the developed method allows one to find out whether the 
emitted photons are twisted without going into details of the process.
This method can be readily extended to the evaluation of the ``twistedness''
of other particles.
%
%
%
%
%
%
%
%
\section*{ACKNOWLEDGEMENTS}
%
%
This work was supported %
by RFBR (Grants No. 16-02-00334 and No. 16-02-00538),
by SPbSU-DFG (Grants No. 11.65.41.2017 and No. STO 346/5-1), and 
by the grant of the President of the Russian Federation (Grant No. MK-4468.2018.2).
%
%
%
%

%
\end {document}